\documentclass[aps,prc,twocolumn,showpacs,floatfix,nofootinbib,preprintnumbers,superscriptaddress,amsmath,amssymb]{revtex4-1}

\usepackage{epsfig}
\usepackage{graphicx}
\usepackage{stmaryrd}
\usepackage{amssymb,tabularx,dcolumn}
\usepackage{supertabular,ltxtable}
\usepackage{longtable}
\usepackage{amsmath}
\usepackage{float}
\usepackage{color}
\usepackage{bm}
\usepackage{CJKutf8}
\usepackage{hyperref}

\renewcommand{\vec}[1]{\mbox{\boldmath $#1$}}

\begin{document}

\begin{CJK*}{UTF8}{gbsn}
\title{The puzzling two-proton decay of $^{67}$Kr}

\author{S.M. Wang (王思敏)}
\affiliation{FRIB/NSCL Laboratory, Michigan State University, East Lansing, Michigan 48824, USA}
\author{W. Nazarewicz}
\affiliation{Department of Physics and Astronomy and FRIB Laboratory, Michigan State University, East Lansing, Michigan 48824, USA}

\date{\today}

\begin{abstract}
Two-proton (2$p$) radioactivity is a rare decay mode found in a few proton-unbound nuclei. The $2p$-decay lifetime and properties of emitted protons carry invaluable  information on nuclear structure in the presence of low-lying proton continuum. The recently measured  $2p$ decay  of $^{67}$Kr  \cite{Goigoux2016} turned out to be unexpectedly fast. Since $^{67}$Kr is expected to be a deformed system, we investigate the impact of deformation  effects on  the $2p$ radioactivity. We apply the recently developed  Gamow coupled-channel framework, which allows for a precise description of three-body systems in the presence of rotational and vibrational couplings. This is the first application  of a three-body approach to a two-nucleon decay from a deformed nucleus. We show that  deformation couplings significantly increase the $2p$ decay width of $^{67}$Kr; this finding explains the puzzling experimental data. The calculated angular proton-proton correlations reflect
a competition between $1p$ and $2p$ decay modes in this nucleus.
\end{abstract}

\maketitle
\end{CJK*}

{\it Introduction}.--
There are very few proton-unbound even-$Z$ nuclei that can  decay  by emitting two protons from their ground states. In such cases, the emission of a single proton is energetically forbidden or strongly suppressed by proton pairing~\cite{GOLDANSKY1960,Pfutzner2004,Blank2008,Pfutzner12,Pfutzner13,Giovinazzo2013,Olsen13}. 
The corresponding half-lives are long enough to characterize this phenomenon as $2p$ radioactivity.
Experimentally, 2$p$ emission from the nuclear ground state (g.s.) was observed for the first time in $^{45}$Fe~\cite{Giovinazzo2002,Pfutzner2002}, and, later on, in $^{19}$Mg~\cite{Mukha2007}, $^{48}$Ni~\cite{Dossat2005,Pomorski2011,Pomorski2014}, and $^{54}$Zn~\cite{Blank2005,Ascher2011}. Interest  in this exotic phenomenon  has been envigorated by measurements of proton-proton correlations in the decay of $^{45}$Fe~\cite{Miernik2007}, $^{19}$Mg~\cite{Mukha2008}, and
$^{48}$Ni~\cite{Pomorski2014}, which have demonstrated the  unique three-body features   of the process and -- when it comes to theory --    the sensitivity  of predictions to the angular momentum decomposition of the $2p$ wave function. The high-quality $2p$ decay data have called  for the development of comprehensive  theoretical approaches, capable of simultaneous description of structural and reaction aspects of the problem~\cite{Blank2008,Pfutzner12}.

The main challenge for theoretical studies of $2p$ radioactivity lies in the model's ability to tackle
 simultaneously  nuclear structure aspects in the internal region  and   the  three-body behavior in the asymptotic region. This becomes especially challenging  for 2$p$ decay since the Coulomb barrier strongly suppresses the wave function at large distances, which also makes the 2$p$ lifetime quite sensitive to the low-$\ell$ wave function components inside the nucleus. So far, most of the theoretical models of $2p$ radioactivity divide the coordinate space into internal and asymptotic regions, where one can use the WKB approach~\cite{GONCALVES2017,Ormand97,Nazarewicz1996}, $R$-matrix theory~\cite{Barker2001,Brown2003}, and current  expression~\cite{Grigorenko2009,Grigorenko2017} to estimate the partial 2$p$ decay width. In our previous work \cite{Wang2017}, we 
 introduced the Gamow coupled-channel (GCC) framework. 
 By utilizing the Berggren-ensemble expansion technique, the GCC model is capable of capturing structure and decay facets of three-cluster systems. Consequently, this tool is very suitable for unraveling the intriguing  features of  $2p$ g.s. decay of $^{67}$Kr.

Being the  heaviest g.s. 2$p$ emitter observed so far, $^{67}$Kr is of particular interest, since it provides unique structural data on medium-mass unbound systems in the presence of  collective modes. The measured 2$p$ decay energy is 1690 $\pm$ 17 keV and the  partial 2$p$ lifetime 20 $\pm$ 11 ms~\cite{Goigoux2016}  is significantly lower than the original  theoretical prediction~\cite{Grigorenko2003}. 
As suggested in Ref.~ \cite{Goigoux2016}, this may be due to 
configuration mixing effects and/or deformation in the daughter nucleus $^{65}$Se. An alternative explanation involves the competition between two-body and three-body  decay channels~\cite{Grigorenko2017}: the partial 2$p$ lifetime can be reproduced only if the two valence protons primarily occupy the $2p_{3/2}$ shell that is supposed to be already filled by the core nucleons. 

The objective of this work is to incorporate a deformed, or vibrational,  core into the GCC model, and study the 2$p$ decay   as the quadrupole coupling  evolves. To benchmark the GCC Hamiltonian, we first consider the simpler  case  of spherical  $^{48}$Ni. Thereafter, we investigate  deformation and configuration mixing effects on  the 2$p$ decay of $^{67}$Kr.

{\it Theoretical framework}--
To describe $2p$ emission, we extend the previously introduced~\cite{Wang2017} three-body core+nucleon+nucleon Gamow coupled-channel (GCC) approach by considering core excitations.
To this end, the  wave function of the parent nucleus is written as 
$\Psi ^{J\pi} =  \left[ \Phi ^{J_p\pi_p} \otimes \phi^{j_c\pi_c} \right]^{J\pi}$,
where $\Phi ^{J_p\pi_p}$ and $\phi^{j_c\pi_c}$ are the wave functions of the two valence protons and the core, respectively.  $\Phi ^{J_p\pi_p}$ is constructed in Jacobi coordinates with the hyperspherical harmonics expansion, of which the hyperradial part $\psi_{\gamma K}(\rho)$ is expanded  in the Berggren basis that includes bound, decaying, and scattering states~\cite{Berggren1968,Wang2017}. $K$ is the hyperspherical quantum number and $\gamma = \{s_1,s_2,S_{12},S,\ell_x,\ell_y,L,J_p,j_c\}$.

The core+$p$+$p$ Hamiltonian of GCC is
\begin{equation}\label{Hcpp}
\hat{H} = \sum^3_{i=c,p_1,p_2}\frac{ \hat{\vec{p}}^2_i}{2 m_i} +\sum^3_{i>j=1} V_{ij}(\vec{r}_{ij}) +\hat{H_c}-\hat{ T}_{\rm c.m.},
\end{equation}
where $V_{ij}$ is the interaction between clusters $i$ and $j$, $\hat{H_c}$ is the core Hamiltonian represented by excitation energies of the core $E^{j_c\pi_c}$, and  $\hat{T}_{\rm c.m.}$ stands for  the center-of-mass term. In this work, the proton-core interaction  $V_{pc}$ is approximated by a Woods-Saxon (WS) average potential including central, spin-orbit and Coulomb terms. At small shape deformations, we applied the vibrational coupling as in Refs.~\cite{HAGINO99,Hagino2001}. At large quadrupole deformations we consider rotational coupling, which was incorporated as in the non-adiabatic approach to deformed proton emitters~\cite{Barmore2000,Kruppa2004}. 

In order to deal with the antisymmetrization between core and valence protons, one needs to eliminate the Pauli-forbidden states occupied by the core nucleons. Due to the fact that the cluster-orbital-shell-model (COSM) coordinates of the valence protons  differ from  Jacobi coordinates, the standard projection technique~\cite{Wang2017} can introduce small numerical errors in the asymptotic region where the wave function is strongly suppressed by the  Coulomb barrier. 
Since the wave function needs to be treated very precisely at large distances, we have implemented the supersymmetric transformation method~\cite{Thompson2000,THOMPSON2004,Descouvemont2003} which introduces an auxiliary repulsive ``Pauli core'' in the original core-$p$ interaction to eliminate  Pauli-forbidden states. For  simplicity, in this work we only project out those spherical orbitals which correspond to the deformed levels occupied in the daughter nucleus.

By using the Berggren basis, the inner and asymptotic regions of the Schr\"{o}dinger equation can be treated on the same footing, and this provides the natural connection between nuclear shell structure and reaction aspects of the problem. The resulting complex eigenvalues  contain information about resonance's energies and decay widths. However, for medium-mass nuclei, due to  the large Coulomb barrier,  proton decay widths are usually below  the numerical precision of calculations ($\sim$10$^{-14}$\,MeV). Still, one can  estimate decay widths through the current expression \cite{Humblet} as  demonstrated in  previous work~\cite{Grigorenko2000,Grigorenko2007,Wang2017}. According to the $R$-matrix theory,  if   the contribution from the off-diagonal part of the Coulomb interaction in the asymptotic region is neglected, the hyperradial wave function of the resonance $\psi_{\gamma K}(\rho)$ is proportional to the outgoing Coulomb function $H^+_{K+3/2}(\eta_{\gamma K},k_p\rho)$~\cite{Grigorenko2009}, where $k_p = \sqrt{2m(E-E^{j_c\pi_c})}/\hbar$ is the complex momentum, $\eta_{\gamma K} = m e^2 Z_{\gamma K, \gamma K}/(k_p\hbar^2)$, and $Z_{\gamma^\prime K^\prime, \gamma K}$ is an effective charge~\cite{DESCOUVEMONT2006,Vasilevsky2001}.
By assuming small Im$(E)$  and adopting the expression  $\psi ^{\prime}/ \psi = k_p {H^+}^\prime/H^+$~\cite{Barmore2000,Kruppa2004}, one can bypass the numerical derivative of the small wave function in the asymptotic region that appears in the original current expression and increase numerical precision dramatically~\cite{Esbensen2000}.

According to Refs.~\cite{Grigorenko2007,Grigorenko2009_2}, the high-$K$ space  of hyperspherical quantum numbers also has some influence on the decay width. 
Since practical calculations must involve some $K$-space truncation,  we adopt the so-called Feshbach reduction  method proposed in Refs.~\cite{Grigorenko2007,Grigorenko2009_2}. This is an adiabatic approximation that allows one to evaluate  the contributions to the interaction matrix elements originating from  the excluded model space.

{\it Hamiltonian and model space} --
For the  nuclear two-body interaction between valence protons we took the finite-range Minnesota force with the original parameters of Ref.~\cite{Thompson1977}. The proton-proton  interaction  has been augmented by the two-body Coulomb force. The core-valence potential 
contains  central, spin-orbit and Coulomb terms. The nuclear average potential has  been taken in a WS form including the spherical spin-orbit term 
with the ``universal'' parameter set \cite{Cwiok1987}, which has been successfully applied to  nuclei from the light Kr region \cite{Nazarewicz1985}. 
The  depth of the WS potential has always been readjusted to the experimental value of $Q_{2p}$.
%
The Coulomb core-proton potential is assumed
to be that of the  charge $Z_ce$ uniformly distributed inside the deformed nuclear
surface ~\cite{Cwiok1987}.

Since $^{48}$Ni is doubly-magic, to discuss its $2p$ decay we limited our calculations to the spherical case. For $^{67}$Kr,  we assumed  a deformed core of  $^{65}$Se 
described by the quadrupole deformation $\beta_2$,
with the unpaired neutron  treated as a spectator. 
According to  calculations~\cite{Aboussir1995,massexplorer,MOLLER2016}, the  $^{65}$Se core has an oblate shape.
Based on the data from the  mirror nucleus $^{65}$Ga~\cite{NNDC}, we assume the g.s. of
$^{65}$Se  to have $J^\pi=3/2^-$~\cite{BROWNE2010} and its rotational (vibrational) excitation
to be a $J^\pi=7/2^-$ state at 1.0758\,MeV.  This estimate is consistent with excitation energies of
$2^+_1$ states in  the neighboring nuclei $^{64}$Zn and $^{66}$Ge~\cite{NNDC}. In our coupled channel calculations, we included  collective states of $^{65}$Se with $J\le j_c^{\rm max} = 15/2^-$; such a choice guarantees  stability of our results.
In particular, we checked that the calculated half-life differs by less than 3\% when varying $j_c^{\rm max}$ from 11/2 to 15/2.

The calculations have been carried out in the model space of $\max(\ell_{x}, \ell_{y})\le 7$ with the maximal hyperspherical quantum number $K_{\rm max}$ = 50 and the Feshbach reduction  quantum number $K_{f}$ = 20, which is sufficient for all the observables studied~\cite{Wang2017,Grigorenko2007,Grigorenko2009_2}. For the hyperradial part, we used the Berggren basis for the $K \le$ 6 channels and the HO basis for the higher angular momentum channels. The complex-momentum contour of the Berggren basis is defined as:  $k  = 0 \rightarrow 0.3-0.1i \rightarrow 0.5 \rightarrow 4 \rightarrow 8$ (all in fm$^{-1}$), with each segment discretized with 50 points. For the HO basis we took  the oscillator length $b = 1.75$\,fm and  $N_{\rm max} = 60$.

{\it Results}.--
We first investigate the spherical  2$p$ emitter $^{48}$Ni, which has been the subject of numerous theoretical studies~\cite{Grigorenko2001,Dossat2005,ROTUREAU2006,BLANK2009,Delion2013,Nazarewicz1996,GONCALVES2017}.  By assuming the experimental  value of $Q_{2p}$ = 1.310\,MeV we obtain  $T_{1/2}=14$\,ms, which agrees reasonably well with  experiment, $T_{1/2}=8.4^{+12.8}_{-7}$\,ms~\cite{Dossat2005} and $3^{+2.2}_{-1.2}$\,ms~\cite{Pomorski2014}. Moreover, we found that calculations with different sets of WS parameters  result in fairly similar decay widths, which is in accord with the conclusion of Ref.~\cite{Nazarewicz1996} that -- as long as the sequence of s.p. levels does not change --  the $2p$ lifetime should  rather weakly depend on the details of the core-proton potential as the tunneling motion of the $2p$ system is primarily governed by the Coulomb interaction.

\begin{figure}[htb]
\includegraphics[width=0.9\linewidth]{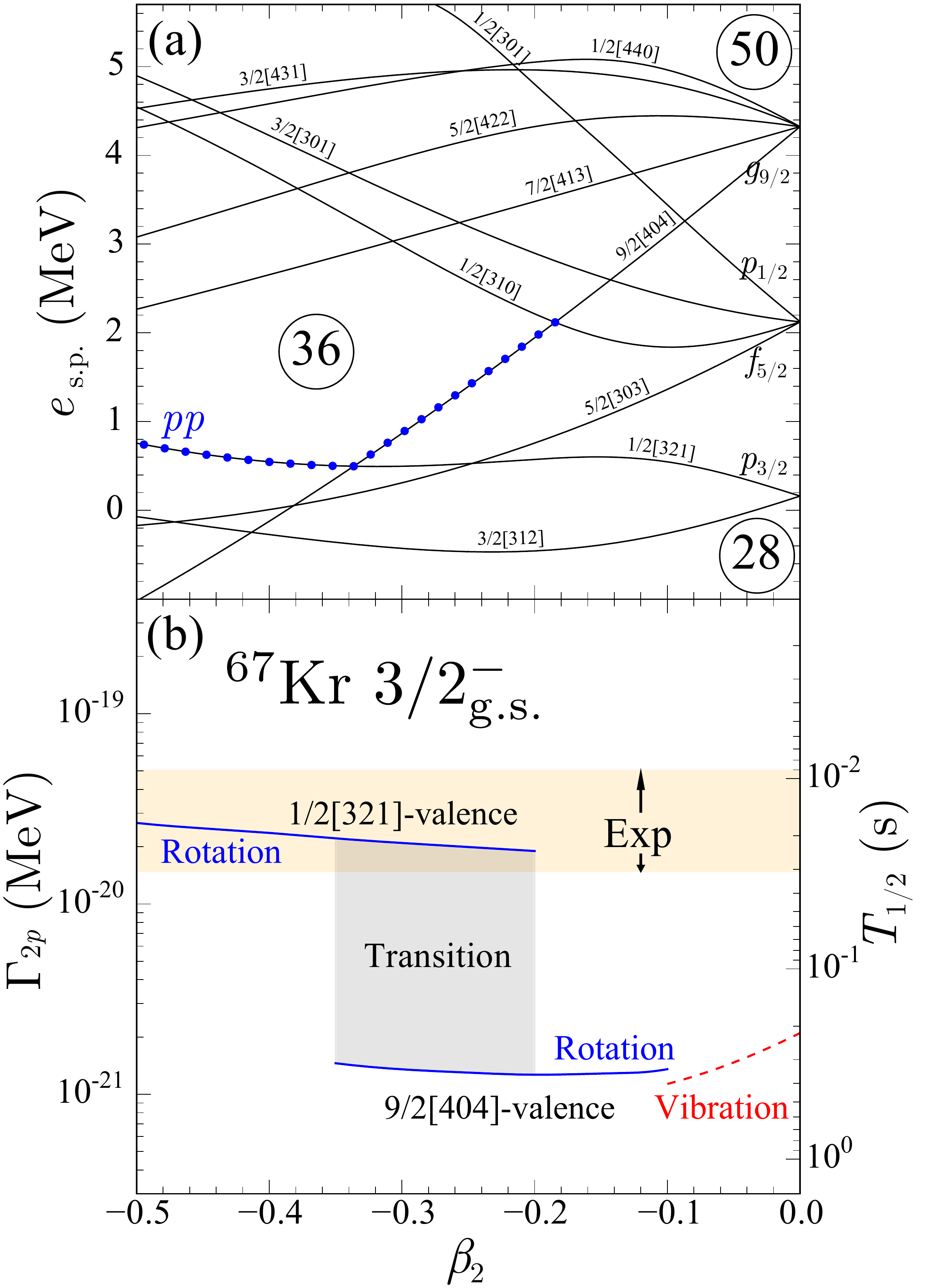}
\caption{\label{Gamma_beta} Top:  Nilsson levels $\Omega$[Nn$_z\Lambda$] of the deformed core-$p$ potential as functions of the  oblate quadrupole deformation $\beta_2$ of the core.
The dotted line indicates the valence level primarily occupied by the two valence protons.
Bottom: Decay width (half-live)  for  the $2p$ g.s. radioactivity of $^{67}$Kr. The solid and dashed lines  mark, respectively, the results within  the rotational and vibrational coupling. 
The rotational-coupling calculations were carried out by
assuming that the 
1/2[321] orbital is either occupied by the core  (9/2[404]-valence) or valence (1/2[321]-valence)   protons. 
}
\end{figure}

The  lifetime of $^{67}$Kr can be impacted by deformation effects \cite{Goigoux2016}. Indeed, studies of one-proton ($1p$) emitters~\cite{Barmore2000,Kruppa2000,Esbensen2000,Davids2001,Davids2004,Kruppa2004,Hagino2001,Florin2003,Arumugam2007} have demonstrated the impact of rotational and vibrational couplings on
$1p$ half-lives. 
Figure~\ref{Gamma_beta}a shows the proton Nilsson levels (labeled by the asymptotic quantum numbers $\Omega$[Nn$_z\Lambda$]) of the WS core-$p$ potential.
At small deformations,  $|\beta_2| \le 0.1$, the valence protons occupy the $f_{5/2}$ shell. The half-life predicted in the vibrational variant of calculations  is $T_{1/2}>218$\,ms, which exceeds the experimental value by over an order of magnitude, see Fig.~\ref{Gamma_beta}b. This result is consistent with previous theoretical estimates~\cite{Grigorenko2003,GONCALVES2017}.

As the deformation of the core increases, an appreciable oblate gap at $Z=36$ opens up, 
due to the downsloping 9/2[404] Nilsson level originating from the $0g_{9/2}$ shell. This gap
is responsible for oblate g.s. shapes of proton-deficient Kr isotopes \cite{Nazarewicz1985,Yamagami2001,Kaneko2004}. The structure of the valence proton orbital changes from the
9/2[404] ($\ell=4$) state at smaller oblate deformations to the 1/2[321] orbital, which has a large $\ell=1$ component. While the exact crossing point of the 1/2[321] and 9/2[404] levels depends on  details of the core-proton parametrization, the general pattern of Fig.~\ref{Gamma_beta}a 
is robust: one expects a  transition from  the $2p$ wave function dominated by
$\ell=4$ components to $\ell=1$ components as oblate deformation increases. 
Figure~\ref{Gamma_beta}b shows the $2p$ decay width predicted in the two limits of the rotational model:
(i)  the 1/2[321]  level belongs to the core, and the valence protons  primarily occupy the 9/2[404] level; and (ii) the valence protons  primarily occupy the 1/2[321] level. 
In reality, as the core is not rigid, proton pairing is expected to produce the diffused Fermi surface; hence the transition from (i) to (ii) is going to be gradual, as schematically indicated by the shaded area in Fig.~\ref{Gamma_beta}b.  
The decreasing $\ell$ content of the $2p$ wave function results in a dramatic increase of the decay width. At the deformation  $\beta_2 \approx -0.3$, which is consistent with estimates from mirror nuclei \cite{Pritychenko2016} and various calculations \cite{Aboussir1995,massexplorer,Nazarewicz1985,MOLLER2016,Pritychenko2016} the calculated  $2p$ g.s. half-live of $^{67}$Kr is 24 ms, which agrees  with experiment~\cite{Goigoux2016}.

\begin{figure}[htb]
\includegraphics[width=0.9\linewidth]{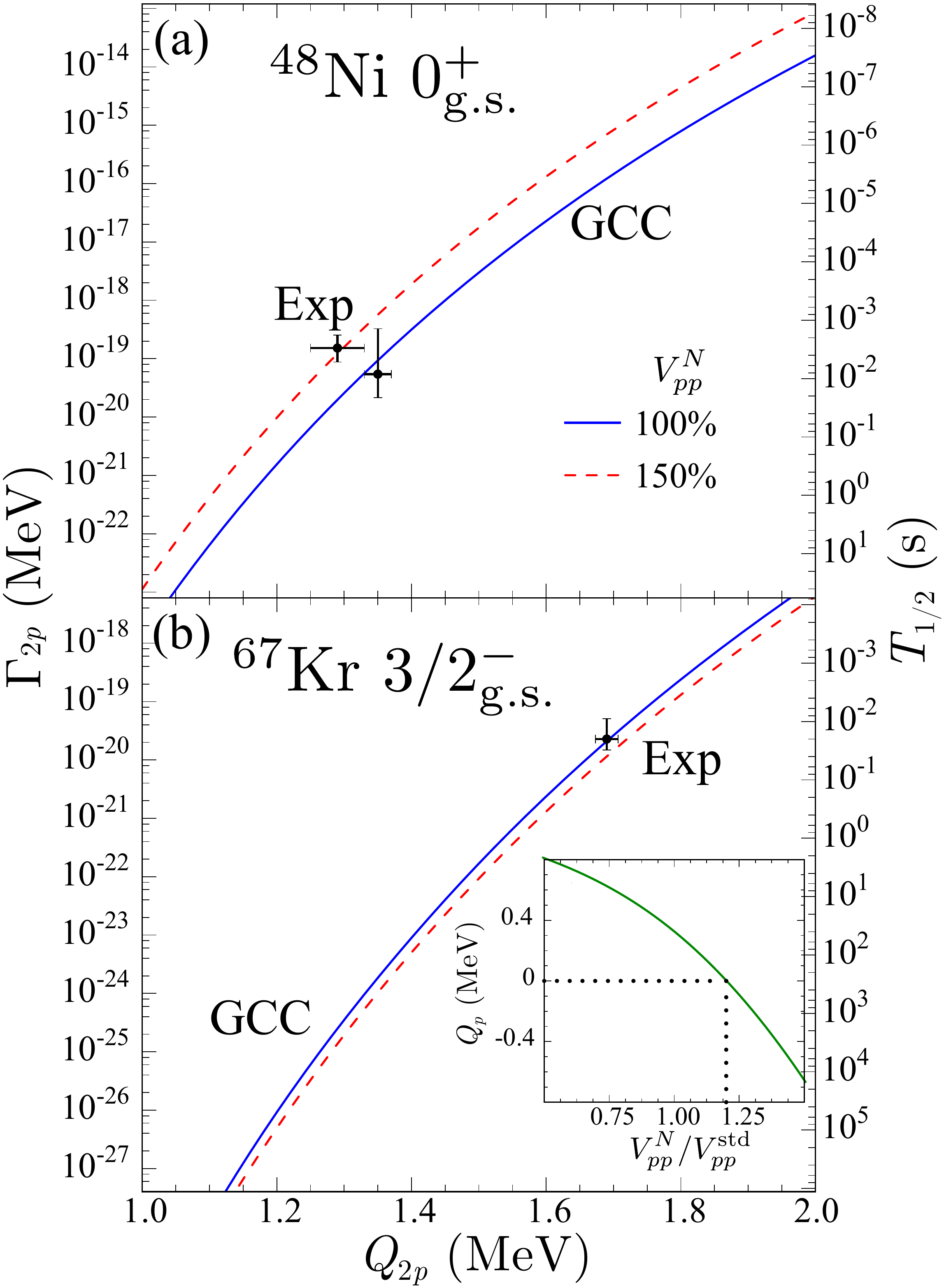}
\caption{\label{Gamma_E} Calculated $2p$ partial width (half-life) of the g.s. decay  of (a) $^{48}$Ni and (b) $^{67}$Kr as a function of $Q_{2p}$. The results obtained with 100\% (solid line) and 150\% (dashed line) strength of the Minnesota force $V^N_{pp}$ are marked. The experimental data are taken from Refs.~\cite{Dossat2005,Pomorski2014} ($^{48}$Ni) and \cite{Goigoux2016} ($^{67}$Kr). The inset  in (b) shows the $1p$ decay energy $Q_{p}$ of $^{67}$Kr at the experimental value of $Q_{2p}$ obtained with  different strengths of $V^N_{pp}$ relative to the original  value  $V^{\rm std}_{pp}$. The $Q_{p}=0$ threshold is indicated by a dotted line.
}
\end{figure}

Since the Minnesota force used here is an effective interaction that is likely to be affected by in-medium effects, one may ask how changes in the proton-proton interaction may  affect the $2p$ decay process. Figure~\ref{Gamma_E} displays  the partial $2p$ width for the g.s.  decay 
of $^{48}$Ni and $^{67}$Kr for two strengths of the $pp$ interaction $V^N_{pp}$. 
The predicted $\Gamma_{2p}$  of $^{48}$Ni is quite sensitive to the strength of $V^N_{pp}$; namely, it increases by an order of magnitude   when the interaction strength  increases by 50\%. 
For the original Minnesota interaction, the $Q_p$ of $^{47}$Co is 1.448\,MeV, i.e.,  
the $1p$ decay channel in $^{48}$Ni is closed. Consequently, further increases in the valence proton interaction strength can only affect the pairing scattering from the $0f_{7/2}$ resonant shell into the low-$\ell$ proton continuum. The corresponding  increase of low-$\ell$ strength in the $2p$ wave function results in  the reduction of half-life seen in
Fig.~\ref{Gamma_E}a. 

The case of  $^{67}$Kr is presented in Fig.~\ref{Gamma_E}b. Here the trend is opposite:
the decay width actually decreases with the strength of  $V^N_{pp}$. To understand this we note that the $1p$ decay channel of the $^{67}$Kr g.s. is open ($Q_p>0)$ for a large range of   interaction strengths, see the insert in Fig.~\ref{Gamma_E}b. At the  standard strength of  $V^{\rm std}_{pp}$,
the predicted $Q_p$ of $^{66}$Br is 1.363 MeV, i.e., 
one expects to see a competition between the sequential and three-body decay in this case.
With the increasing pairing strength, the odd-even binding energy difference grows, and the $1p$ channel gets closed around  $V^N_{pp}/V^{\rm std}_{pp}=1.2$. The further increase of $V^N_{pp}$ strength results in pairing scattering to higher-lying proton states originating from $0g_{9/2}$ and $0f_{5/2}$ shells with higher $\ell$ content, see Fig.~\ref{Gamma_beta}. Both effects explain the reduction of $\Gamma_{2p}$ seen in Fig.~\ref{Gamma_E}b.

\begin{figure}[htb]
\includegraphics[width=0.8\linewidth]{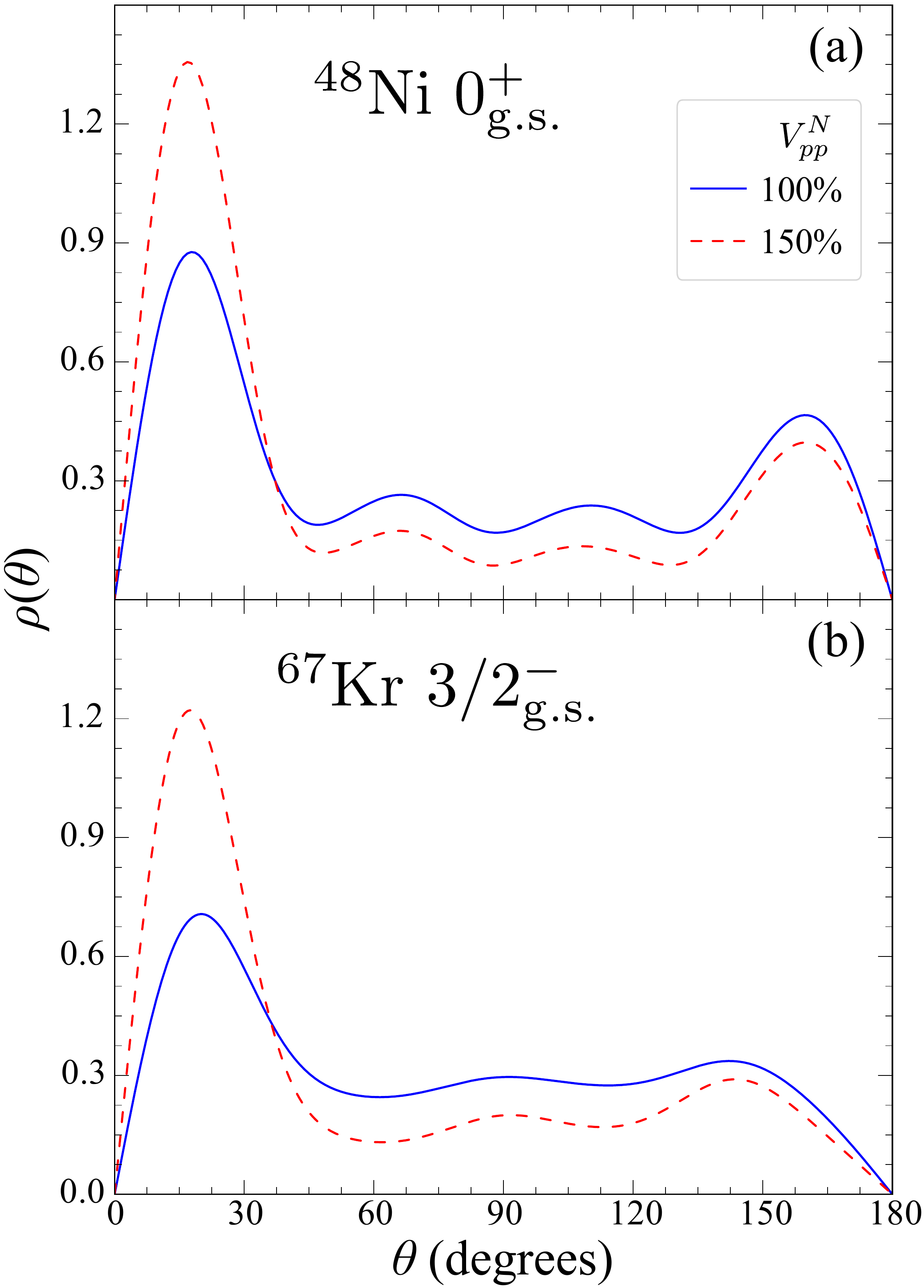}
\caption{\label{ang_cor} Two-proton angular correlation for the g.s. of (a) $^{48}$Ni and (b) $^{67}$Kr 
obtained with the Minnesota force of standard strength (solid line) and 50\%-increased strength  (dashed line). 
}
\end{figure}
Since the $1p$ channel is most likely open for $^{67}$Kr~\cite{Grigorenko2017}, it is interesting to ask: How large is the diproton component in the $^{67}$Kr decay?
To this end, in Fig.~\ref{ang_cor} we study the 2$p$ angular correlations~\cite{Papadimitriou11,Wang2017}
for the g.s. decays of $^{48}$Ni and $^{67}$Kr. 
In both cases,  a diproton-like structure corresponding to a peak 
at  small opening angles  is very pronounced. Interestingly, according to our calculations, the two valence protons form very similar configurations in $^{48}$Ni and $^{67}$Kr. Namely, for $^{48}$Ni the dominant $(S_{12},\ell_x,\ell_y)$ configurations in T-type Jacobi-coordinate are 58\% (0, 0, 0) and 30\% (1, 1, 1), while the corresponding amplitudes for   $^{67}$Kr  are 59\% and 27\%.  The diproton peak in $^{67}$Kr is slightly lower than that in $^{48}$Ni due to the fact that sequential decay is energetically allowed in $^{67}$Kr. 
The $1p$  decay width of $^{67}$Kr estimated by the core-proton model is 8.6$\times 10^{-20}$ MeV, which has the same order of magnitude with the 2$p$ decay width. Consequently, the 2$p$ decay 
branch in $^{67}$Kr is expected to compete with the sequential decay. With the  pairing strength increased by 50\%  the diproton peak in $\rho(\theta)$ becomes strongly enhanced, see Fig.~\ref{ang_cor}, as the $1p$ channel gets closed.

{\it Conclusions}.--
We extended the Gamow coupled-channel approach by introducing couplings to  core excitations. 
We demonstrated that  deformation effects are important for  the 2$p$ g.s. decay of $^{67}$Kr.
Due to the oblate-deformed  $Z=36$  subshell at  $\beta_2 \approx-0.3$,
the Nilsson orbit 1/2[321] with large $\ell=1$ amplitude becomes available to valence protons. 
This results in a significant increase of the $2p$ width of $^{67}$Kr, in accordance with experiment.

The sensitivity of $2p$ lifetime to the proton-proton interaction indicates that the pairing between the  valence protons can strongly influence   the decay process. Through the comparison of one-proton decay energies and angular correlations between $^{48}$Ni and $^{67}$Kr, we   conclude that there is a competition between $2p$  and $1p$ decays in $^{67}$Kr, while 
the decay of  $^{48}$Ni has a $2p$ character.

In summary, the puzzling $2p$ decay of $^{67}$Kr has been naturally explained in terms of the shape deformation of the core. The explanation is fairly robust with respect to  the details of the GCC Hamiltonian. We conclude that the Gamow coupled-channel framework provides a comprehensive description of structural and reaction aspects of three body decays of spherical and deformed nuclei.

\begin{acknowledgements}
Discussions with K{\'e}vin Fossez, Futoshi Minato, Nicolas Michel, Jimmy Rotureau, and Furong Xu are acknowledged. We appreciate helpful comments from Zach Matheson. This material is based upon work supported by the U.S.\ Department of Energy, Office of Science, Office of Nuclear Physics under  award numbers DE-SC0013365 (Michigan State University), DE-SC0018083 (NUCLEI SciDAC-4 collaboration), and DE-SC0009971 (CUSTIPEN: China-U.S. Theory Institute for Physics with Exotic Nuclei).
\end{acknowledgements}

\bibliographystyle{apsrev4-1}
\bibliography{2p_decay}
\end{document}